# Impurity-driven turbulence opens a pathway to ELM-free operation and enhanced pedestal stability in tokamaks

Santanu Banerjee[1], T. Macwan[2], A. Bortolon[1], R. Groebner[3], K. Barada[4], R. Maingi[1], T. Osborne[3], T. L. Rhodes[4], C. Chrystal[3] and Z. Yan[5]

[1]Princeton Plasma Physics Laboratory, PO Box 451, Princeton, NJ 08543-0451, USA
[2]Lawrence Livermore National Laboratory, Livermore, CA, 94550 USA
[3]General Atomics, PO Box 85608, San Diego, CA 92186-5608, USA
[4]Department of Physics and Astronomy, UCLA, Los Angeles, CA 90095, USA
[5]University of Wisconsin-Madison, Madison, WI 53706-1687, USA

email: sbanerje@pppl.gov

**Abstract**

Edge-localized modes (ELMs) impose severe transient heat, and particle loads on plasma-facing components, posing a critical challenge for steady-state operation of tokamak fusion reactors. Existing ELM control techniques either rely on externally applied perturbations or operate within narrow parameter windows, raising concerns for reactor scalability. Here we demonstrate that controlled injection of a low-Z impurity can fundamentally modify pedestal transport and stability, enabling access to long ELM-free periods through impurity-driven turbulence. Using boron (B) powder injection in the DIII-D tokamak, we observe a progressive reduction of ELM frequency, culminating in long ELM-free phases. Pedestal stability analysis reveals a pronounced decoupling of peeling and ballooning stability boundaries at moderate B injection levels, opening a stability channel toward super-high confinement operation. At higher injection rates, long (~300 ms) ELM-free periods are achieved. Fluctuation measurements show that boron injection selectively enhances low-frequency pedestal turbulence, increasing inter-ELM particle transport and regulating pedestal gradients. The establishment of a feedback loop between turbulence, particle transport, and the resulting modification of pedestal conditions, indicated by the observed hysteresis loop in the evolution of density fluctuations in response to the B injection rate, is presented.

## 1. Introduction

Global confinement of the tokamak plasma scales with the height of the pressure achieved at the pedestal top [1], where 'pedestal' is referred to the edge transport barrier, spanning through the outer few percentages of High-confinement mode (H-mode) plasmas. The achievable height of the pedestal is limited by semi-periodic Magneto-HydroDynamic (MHD) instabilities caused by both the pressure gradient as well as the current profile, resulting in the so-called Edge Localized Modes (ELMs) [2,3]. ELMs are responsible for huge heat and particle loads on the plasma facing components, mainly divertors and thereby limiting their lifetime. Active research is ongoing for deliberately pacing, mitigating, or suppressing ELMs either by external means like pellet injection [4,5] and/or resonant magnetic perturbations [6,7], or by development of natural no-ELM scenarios like quiescent H-mode (QH) [8,9], wide pedestal QH (WPQH) [10], enhanced D-alpha (EDA) H-mode [11,12], improved (I) -mode [13,14] etc. The external means are often non-trivial to implement, especially in a reactor grade machine. Further, all these natural no-ELM operating scenarios are achievable within narrow operational windows, in terms of either collisionality, $q_{95}$, heating power, rotation and rotation shear, density at the pedestal top etc. or a combination of these parameters [15]. Hence, the scalability of these scenarios to reactor grade operations is still uncertain. Apart from these scenarios, ELM-free operations are achieved sometimes, for example, with low-Z impurity injection like Li [16]. However, ELM-free phases are mostly transient and may suffer from problems like impurity buildup in the core, difficult to control density rise, core MHD, large ELM-bursts at the end of the ELM-free phase or H-L back transition [16-18].

There are other proven ways of modulating the ELM frequency ($f_{ELM}$) via turbulence driven transport at the maximum gradient region of the pedestal. One such route could be by varying the heating mix between electron cyclotron heating (ECH) and neutral beam injection (NBI) or changing the ECH deposition location [19-24]. It has also been demonstrated in DIII-D that $f_{ELM}$ can be varied significantly by simply varying the time interval between the start of auxiliary heating (e.g. NBI) with respect to the time when the plasma current reached the flat top [25]. Notably, in all these approaches, the role of pedestal turbulence is emphasized to modulate the pedestal transport during the inter-ELM period.

As an alternate approach, low-Z impurity is injected, and ELM suppression is achieved in a few tokamaks. Li evaporative coating on wall or Li powder injection has resulted in ELM suppression in NSTX [26-28] and EAST [29,30]. Recently, quasi-steady ELM suppression with boron (B) powder injection has been demonstrated in EAST [31]. Further, a comparison of ELM suppression in similar discharges via either Li or B injection indicated that there might be a difference in the mechanism of ELM suppression between these two different types of low-Z impurities [32].

While complete ELM suppression is not achieved, Li powder injection increased the duration of the enhanced pedestal phases in DIII-D, characterized by increased pedestal pressure and width along with ELM-free periods to up to 350 ms. Li injection also increased the likelihood of a transition to the enhanced phase [16]. Apart from Li injection, improvements in the pedestal pressure and global energy confinement were also observed on DIII-D with both 3 and 6 MW of neutral beam (NBI) heating with neon (Ne) injection [33]. Decrease in $f_{ELM}$ was also reported with Ne injection. B powder injection in the grassy ELM regime of KSTAR resulted in increased ELM amplitude, which was attributed to decreased pedestal collisionality [34]. On the other hand, ELM mitigation was demonstrated in KSTAR with boron nitride injection [35]. Nevertheless, in all these cases some kind of fluctuation mode, like the bursty chirping mode (BCM) in DIII-D [16] or the edge harmonic mode (EHM) in EAST [31], was observed to be responsible for enhanced particle transport at the maximum gradient region of the pedestal.

In this experiment, 76% decrease of $f_{ELM}$ is demonstrated with B powder injection, using the impurity power dropper (IPD) [36,16] in DIII-D. Further, decoupling of the peeling and ballooning stability boundaries in the pedestal stability space is demonstrated for the first time with low-Z impurity injection. The pedestal stability space is represented by the edge current density ($j_{edge}$) and the average gradient of the pedestal pressure ($\alpha = \nabla p^{ped}$). This decoupling can provide easier access to the super-H mode [Snyder NF 2019] in future. Long ELM-free periods (~300 ms) are also demonstrated with even higher B injection rate. However, these ELM-free periods, even though long, cannot be sustained and ended in large ELMs causing ~15% decrease in stored energy following these ELMs.

Since increased turbulence and turbulence driven transport is a generic feature of the pedestal of H-mode plasmas whenever some impurities are injected, like Li, Ne etc., the goal of this paper is to verify if that is also the case for B injection in DIII-D and how that increased turbulence is improving pedestal stability, which in turn is paving the way towards achieving long ELM-free periods of operation. It is also necessary to investigate what features of the pedestal are modified with B injection and if the turbulence scenario is also generic for all injected impurities like Li, Ne or B.

## 2. Results

### *2.1. Progressive reduction of $f_{ELM}$ and achieving long ELM-free periods by B injection*

This experiment is performed with type-I ELMing H-mode plasmas in the lower single null (LSN) scenario in DIII-D [37], heated with 2 MW NBI (injected power, $P_{inj}$). Plasma current $I_p$, toroidal magnetic field $B_T$ and hence $q_{95}$ are kept constant is these discharges at 1 MA, 2T and 5 respectively. The normalized plasma pressure, $\beta_N$ (= $\beta a B_T/I_p$, $a$ being the minor radius and $\beta$ being the ratio of the plasma pressure to the magnetic pressure, expressed in percentage) is ~ 1.5 in all the discharges of this series. These discharges are operated with a total input co-current torque of 2.1 Nm. Plasma discharges in this experiment were designed to study the effect of real time wall conditioning with B powder injection on the plasma conditions. The rate of B injection per discharge, from the IPD, was cautiously increased along the day getting to the higher rates towards the end of the day. The signs of effective wall conditioning became apparent from the fueling and decreased impurity content in the L-mode phase (not shown here). Once injected, the B powder takes ~900 ms to reach the plasma. Here, mostly three representative discharges from the data set will be discussed. Discharge #179863 is the reference discharge without B injection, while discharges #179868 and #179873 represent low and high B injection rates at 2 and 4.5 mg/s respectively. Apart from these, two other discharges with even higher injection rates, will be occasionally discussed in this paper from the point of view of confinement and ELM -free periods of operation. These are discharges #179877 and #179882 with higher B injection rates of 9.7 and 12.7 mg/s respectively.

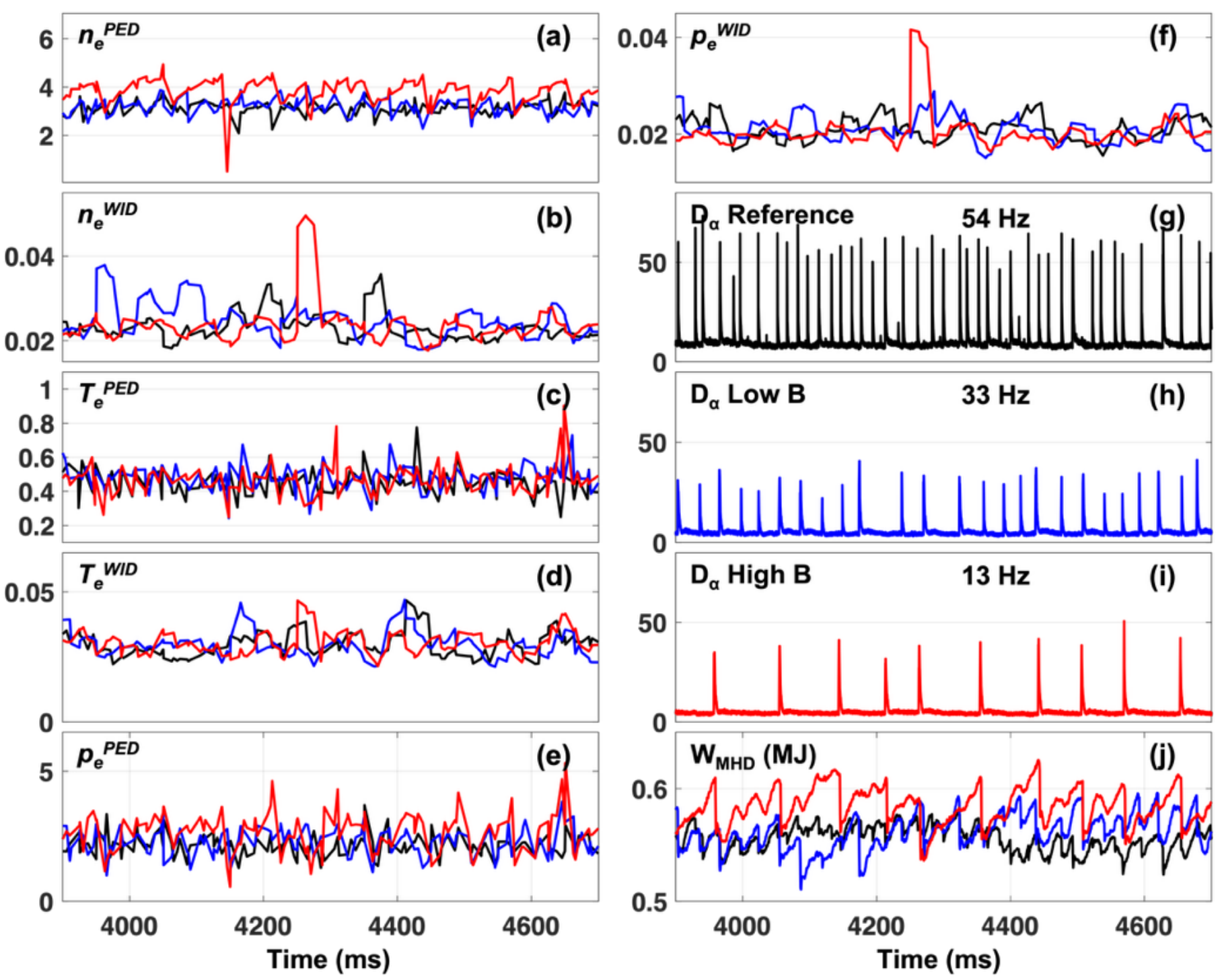

*Fig. 1: (a-f): time evolution of the electron density ($10^{19}$ $m^{-3}$), temperature (keV) and pressure (kPa) at the pedestal top and the respective pedestal widths (m), measured by TS; (g-i) time evolution of the divertor Dα signal (AU) showing the ELM bursts; (j): Stored energy of the plasma. All the time traces are for the reference (black, #179863), low B (blue, #179868) and high B (red, #179873) injection discharges.*

A multi-chord, multi-pulse Thomson Scattering (TS) system is used to measure electron density ($n_e$) and electron temperature ($T_e$) profiles at a high spatial and time resolution, along a vertical chord in the machine [38]. Time evolution of the pedestal parameters like the height and width for $n_e$, $T_e$ and electron pressure ($p_e$) are shown in Fig. 1(a – f). An increase in $n_e^{ped}$ and $p_e^{ped}$ is observed in the inter-ELM periods of the high B discharge (red). $f_{ELM}$ dropped with B injection rate and it is reduced by ~76% for the high B discharge as compared to the reference discharge, as shown in Fig. 1(g – i). Evolution of the total stored energy ($W_{MHD}$) is shown in Fig. l(j). Enhanced $W_{MHD}$ in the inter-ELM periods is observed for the high B discharge.

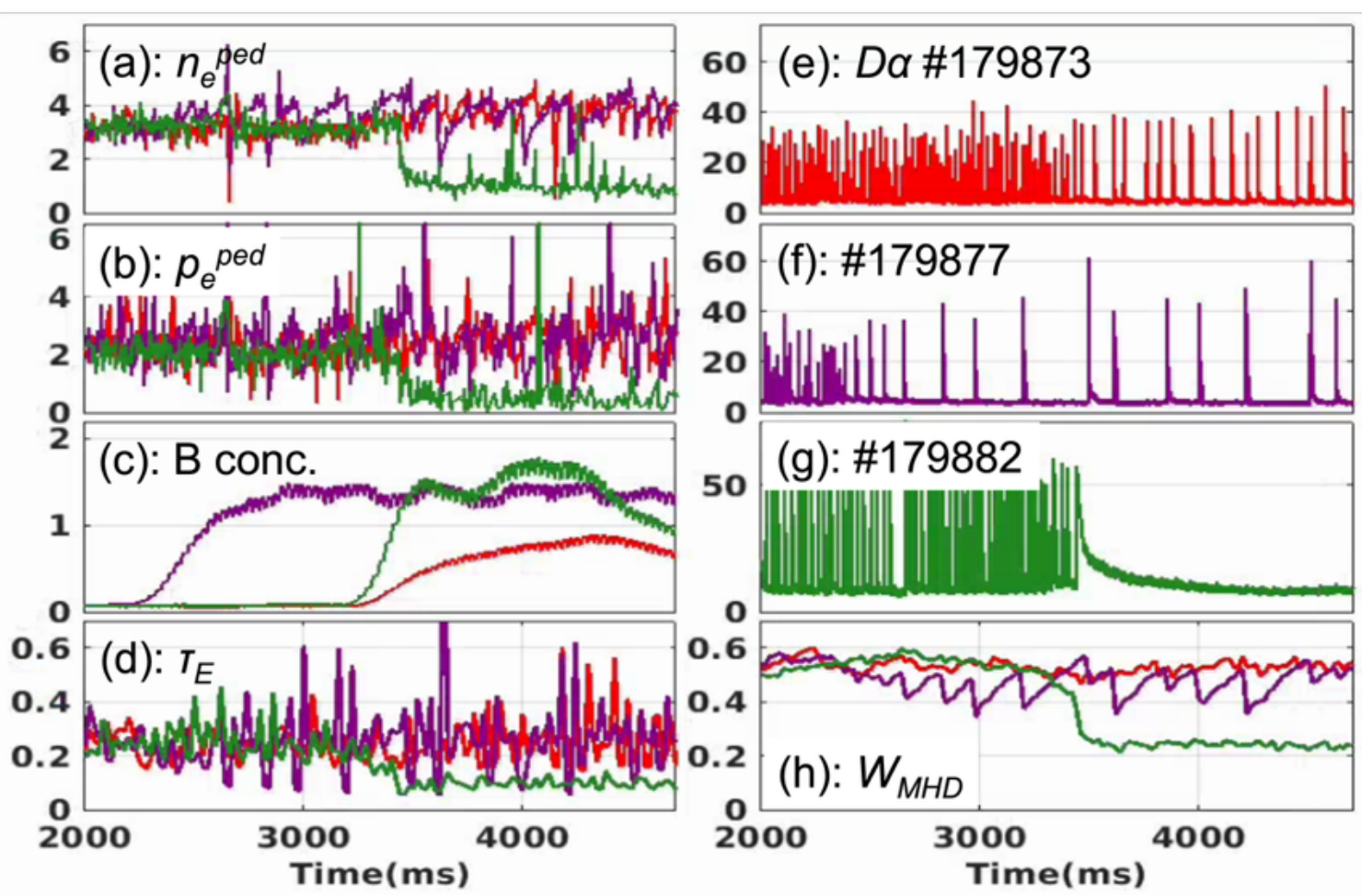

*Fig. 2: (a-b): Time evolution of $n_e^{ped}$ ($10^{19}$ $m^{-3}$) and $p_e^{ped}$ (kPa) for three discharges, like the high B (red, #179873, 4.5 mg/s), even higher B, (dark magenta, #179877, 9.7mg/s) and highest B (green, #179882, 12.7 mg/s); (c): evolution of B in the SPRED spectrometer; Note that the B injection timing is different for these discharges. B is injected at 2.5 s for #179873 and #179882, while it is at 1.5s for #179877; (d) evolution of $\tau_E$ (s); (e-g): evolution of divertor Dα signal showing the ELM bursts; (h) evolution of $W_{MHD}$ (MJ).*

As a result of even higher amount of B injection, discharge #179877 shows long ELM-free periods (~300 ms). After the end of each ELM-free period there are bigger ELM bursts and bigger transient heat flux bursts on the divertor (not shown here) compared to the reference discharge. Stored energy ($W_{MHD}$) just prior to the big ELMs at the end of each ELM-free phase is comparable to the high B discharge (#179873). However, with each big ELM at the end of ELM-free phases there is a ~15% decrease in the stored energy. But there is no significant growth of core MHD activities during the ELM-free periods and the discharge did not experience H-L back transition. Any further increase in the B injection rate resulted in severe confinement loss and subsequently H-L back transition (#179882). Fig. 2 shows the time evolution of the $n_e^{ped}$, $p_e^{ped}$, B concentration measured by the SPRED spectrometer [39], evolution of the energy confinement time $\tau_E$, Dα traces for these two discharges in comparison with discharge #179873 and the evolution of $W_{MHD}$. It is apparent from the trend of $f_{ELM}$ reduction and the ELM-free periods, that a more gradual increase in the B injection rate, with an amount, somewhere between 4.5 (#179873) and 9.7 mg/s (#179877), may provide longer ELM free periods with enhanced pedestal pressure, as observed earlier with Li [16]. Note that there is no significant difference in the level of the energy confinement time ($\tau_E$) between #179873 and #179877, showing that the energy transport might not have been affected even with 9.7 mg/s of B injection rate, while achieving long ELM-free periods. The interplay (if any) of $f_{ELM}$, $W_{MHD}$ and the ELM size is deciphered in the next sub-section.

### *2.2. Relation between $f_{ELM}$ and energy loss per ELM*

In this sub-section we will discuss about the relation between $f_{ELM}$ and energy loss per ELM. Fig. 3(a) shows the ELM-synchronized [25,40,41] evolution of $W_{MHD}$, calculated from the fast EFIT reconstructions [25] in the 3700-4800 ms time window for the three representative discharges. The ELM synchronization is performed by detecting the ELMs in the Dα amplitude beyond a user-defined threshold. In this dataset, ELMs those are larger than ~20% of the largest ELMs in the Dα amplitude are detected and this threshold is identical for the reference as well as the B injected discharges. Fig. 3(a) shows that the variation in $W_{MHD}$ is <4% among these 3 representative discharges, prior to an ELM event. Further, energy loss per ELM ($\Delta W_{MHD}$) is smallest for the reference, no B, discharge and then increases progressively with B concentration. Fig. 3(b) shows the variation of $\Delta W_{MHD}$ as a function of $f_{ELM}$. Like the observations in ASDEX-Upgrade and other tokamaks [Fig. 5 of 40], a negative linear correlation between $\Delta W_{MHD}$ and $f_{ELM}$ is observed for $f_{ELM}$ <100.

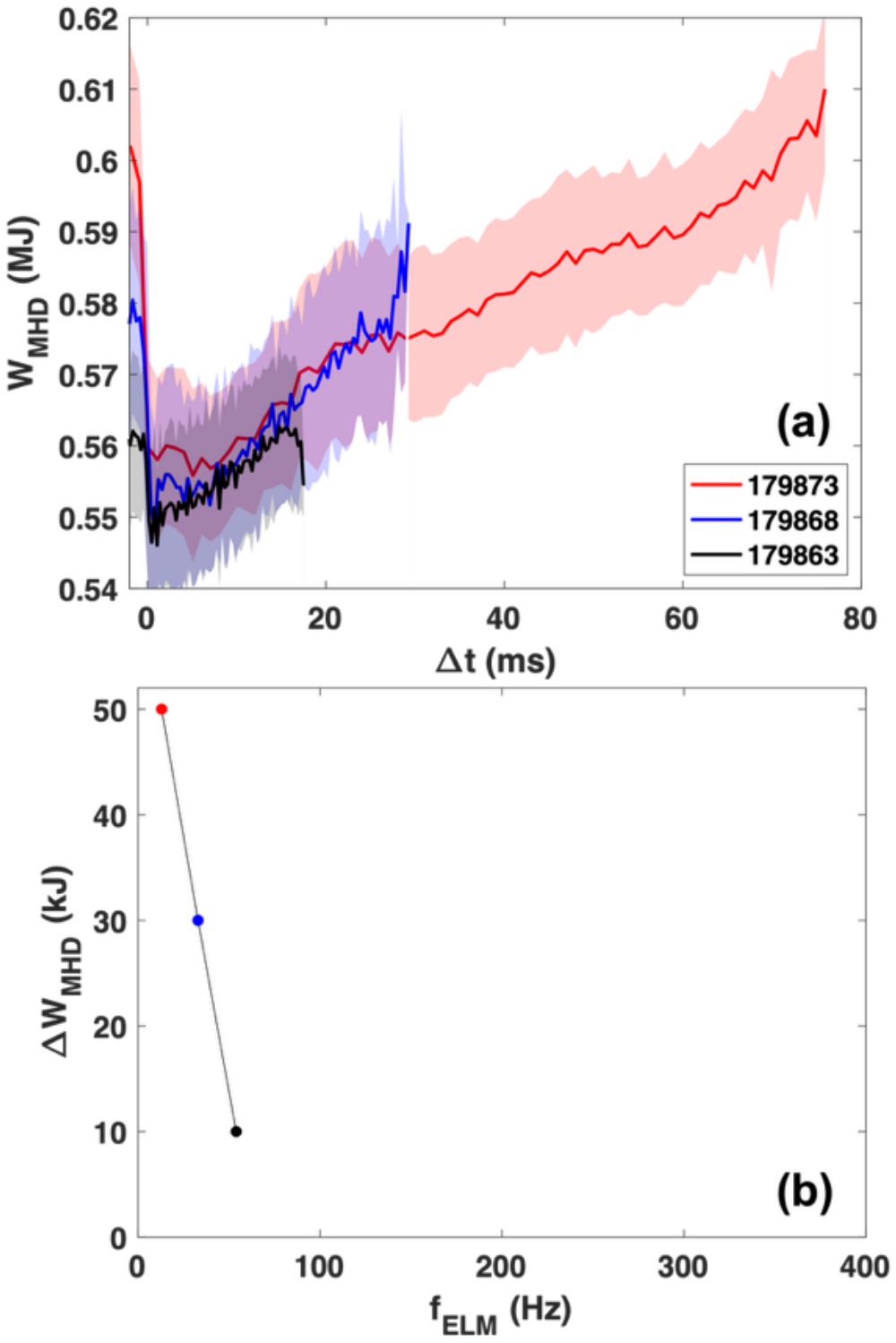

*Fig. 3: (a): ELM-synchronized evolution of $W_{MHD}$ in the inter-ELM period for the three representative discharges. The shaded areas represent the ± standard deviation for all the ELMs in the time window of 3700-4800 ms; (b): Relation between $f_{ELM}$ and $\Delta W_{MHD}$ for these three discharges. Scale kept same as Fig. 5 of [40] for easier comparison.*

### *2.3. Increased pedestal pressure achieved by B injection*

To check for the role of inter-ELM turbulence driven transport and the pedestal stability for ELMs, first we need to analyze the pedestal profiles in the 70-99% of the ELM cycle. Differences (if any) in the saturated pedestal profiles can lead us towards the physical mechanism that B injection might have unveiled and help us to understand the reason behind the observed variation in $f_{ELM}$ better.

Fig. 4(a-c) shows the kinetic fitted profiles of electron density ($n_e^{ped}$), temperature ($T_e^{ped}$), and hence pressure ($p_e^{ped}$) in the pedestal from 70-99% of the inter-ELM period for the three representative discharges. A 300 ms time window is used for this analysis. The negative gradients

are shown with dashed lines. $n_e^{ped}$ increases sharply with high B, but $T_e^{ped}$ initially rises for low B and then drops a little for the high B case. A steady increase in $p_e^{ped}$ is observed with B injection. Consequently, the gradients of $n_e^{ped}$ and $p_e^{ped}$ are significantly higher for the high B case as compared to the reference discharge. However, the gradient of $T_e^{ped}$ is lower. Fig. 4(d-f) shows the kinetic profiles of $n_i$, $T_i$ (solid) and $T_e/T_i$ (broken), total pressure ($p_{tot}$ in solid) and its gradient (broken). A gradual decrease in $n_i$ is observed, as expected, with increasing B. No significant change in $T_i$ is apparent though. Progressive increase in $p_{tot}$ is observed with B. Fig. 4(g-i) shows the kinetic profiles of carbon (C) density ($n_C$, solid) and B density ($n_B$, dotted), bootstrap current ($j_{BS}$) and the effective charge ($Z_{eff}$). Interestingly, $n_C$ increases with $n_B$ and in the high B case $n_B > n_C$. Note that DIII-D being a tokamak with carbon first wall, has carbon as the dominating impurity in general. There is a strong increase in $j_{BS}$ with B. Finally, $Z_{eff}$ is almost doubled in the high B discharge compared to the reference discharge.

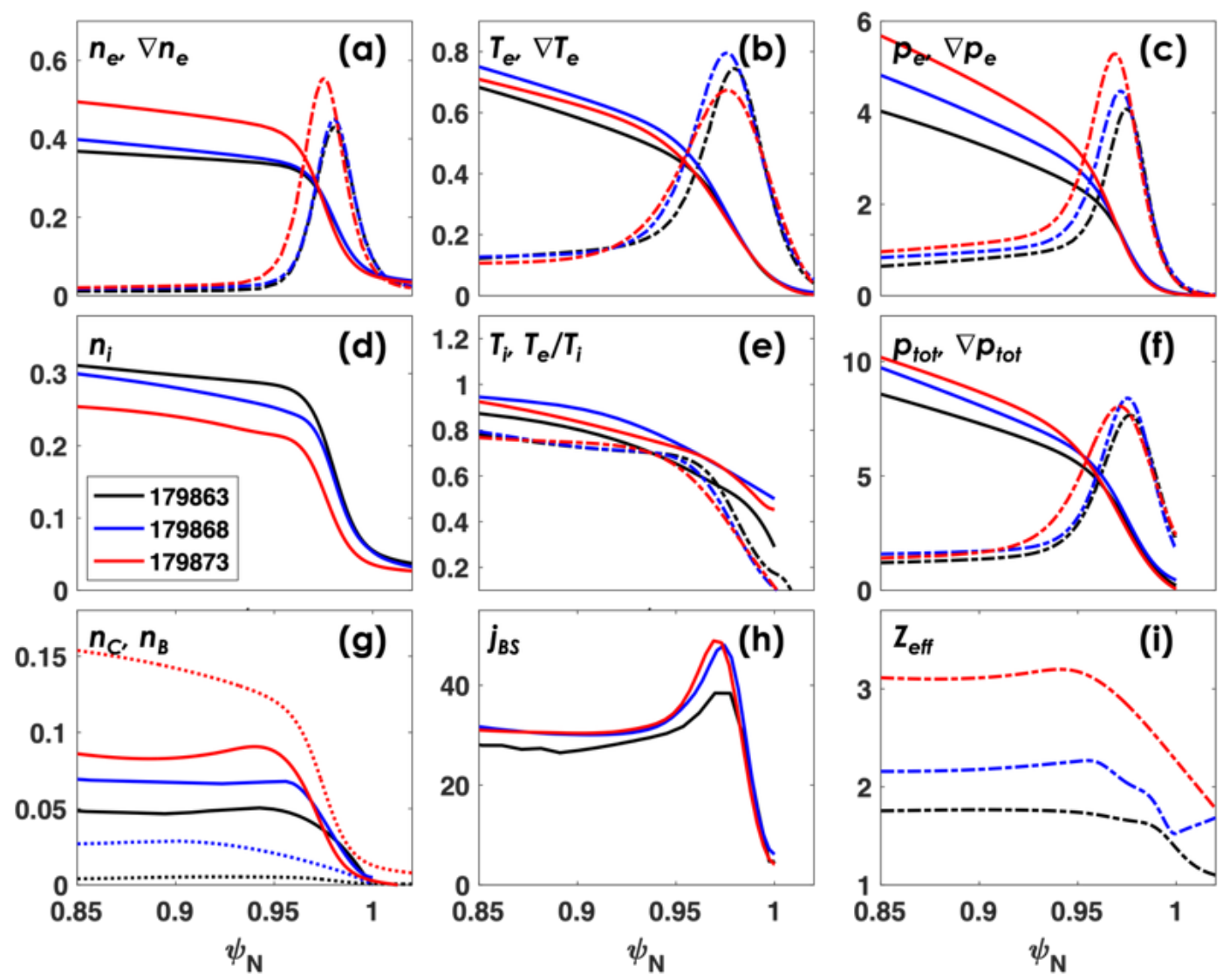

*Fig. 4: (a-c) kinetic profiles of the $n_e$ ($10^{20}$ $m^{-3}$), $T_e$ (keV) and $p_e$ (kPa) are shown at the pedestal. The negative gradients are shown with broken lines; (d-f): kinetic profiles of $n_i$ ($10^{20}$ $m^{-3}$), $T_i$ (keV, solid) and $T_e/T_i$ (broken), total pressure ($p_{tot}$ in solid, kPa) and its gradient (broken); (g-i): kinetic profiles of carbon density ($n_C$, solid, $10^{19}$ $m^{-3}$) and B density ($n_B$, dotted, $10^{19}$ $m^{-3}$), bootstrap current density ($j_{BS}$) and the effective charge ($Z_{eff}$).*

Noted differences in the key pedestal parameters indicate that there could be substantial differences in the pedestal stability as well as the turbulence scenario at the maximum gradient region. Peeling-ballooning (PB) stability of the pedestal will be analyzed next, followed by the observed fluctuations.

### *2.4. Improved ballooning stability achieved leading towards the opening of the super-H channel*

Introduction of impurities can alter the pedestal stability by changing the peeling and/or ballooning stabilizations via altering $Z_{eff}$ (fuel dilution) [16]. Pedestal stability analysis, done with the ELITE [2] code, revealed that the ballooning stability is improved progressively with B injection and the peeling and ballooning thresholds are widely decoupled in the high B discharge. Fig. 5(a-c) shows the PB stability for the three representative discharges in the quasi-stationary phase. For the discharges with B injection, the analysis window is chosen following the injection and when the effect of B is apparent and stationary. ELITE is a MHD stability (eigenvalue) code. It is a single fluid code which includes diamagnetic stabilization via analytic models or fits to two fluid calculations. The stability boundary is determined by the ratio of growth rate to diamagnetic stabilization term on the grid of the normalized pedestal pressure gradient ($\alpha \propto dp^{PED}/d\rho$, $\rho$ being the normalized plasma radius) and normalized pedestal current density ($j_N$), and then tracing the contour for which this ratio is 1 [16,25,42]. Note that the form of the diamagnetic stabilization derived from BOUT++ calculations, as discussed in [42], is used in this paper. However, we have adjusted the critical value of the toroidal mode number ($n_{crit}$), where the transition point for the shift to weaker diamagnetic stabilization occurs in the bilinear fit (refer to Fig. 4 in [42]). We have used $n_{crit} = 21.7/q_{alfmax}$, where $q_{alfmax}$ is the $q$ at peak alpha value to fit the experiment, based on the approximate value used in [42]. This BOUT++ model best fits the experimental data from DIII-D and has been used for a wide range of plasma configurations and operation regimes. This adjustment seems necessary as the stable (blue) region in the $j$-$\alpha$ diagram, in each panel, extends to the purple region towards the right and that transition is somewhat similar for all the B level cases, irrespective of the shape of the peeling and ballooning boundaries. This suggests a relatively weaker diamagnetic stabilization for all these discharges in the series compared to usual H-mode discharges in DIII-D. Note that the ELITE results presented here are also tested for high toroidal mode number (high-$n$, $n \geq 60$) modes and the stability boundaries remain similar. Nevertheless, the stability channel opened with high B injection, due to the decoupling can provide the opportunity to achieve much higher pedestal pressure, like the Super H-mode regime [43,44], if the channel is carefully navigated through. Such decoupling of the PB boundaries and opening an access channel to the second stability regime was also observed during ELM-free periods in MAST-U recently [45]. This is for the first time it is demonstrated that the stability channel for accessing super-H mode can be opened up by impurity injection. The shape of these discharges is shown in Fig. 5(d). There is a small variation between the reference and B injected discharges. The outer strike point is on the floor for the reference discharge while it is on the shelf for the B injected discharges. Note that, progressive reduction of $f_{ELM}$ is observed with B despite this small variation in shape in the reference discharge and hence, it seems to be insignificant in the context of this study. In fact, the bottom triangularity ($\delta_{bot}$) of the reference discharge is much higher ($\delta_{bot} = 0.72$) as compared to that of the B injected discharges ($\delta_{bot} = 0.61$). It has shown earlier that increased triangularity leads to lower $f_{ELM}$ and larger energy loss per ELM [20,40].

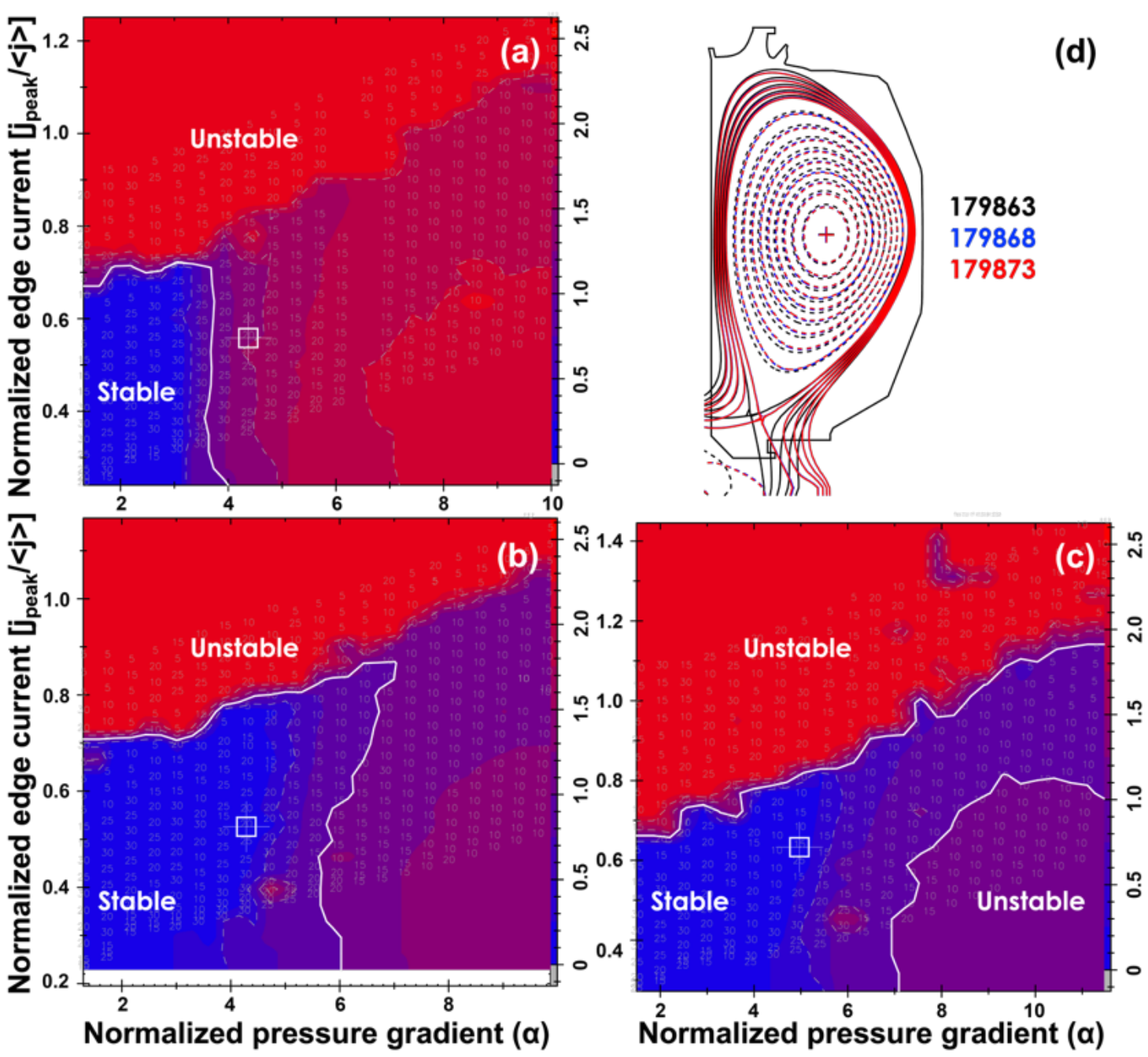

*Fig. 5: (a-c): PB stability of the reference (#179863), low B (#179868) and the high B (#179873) discharges. Operating point of each discharge is shown by a square; (d): Shape of the three discharges are shown from EFIT reconstructions.*

It is worthwhile here to revisit the role of diamagnetic drifts in the pedestal stability against ELMs. The total diamagnetic drift creates a stabilizing effect against certain MHD instabilities like peeling-ballooning modes (the drivers of ELMs) and kinetic ballooning modes (KBMs) [46,47]. The stabilization effect of diamagnetic drifts is particularly important for high-*n* ballooning modes. This expands the stable region for the pedestal pressure gradient, allowing for higher pressure before the plasma becomes unstable. A shift in the stability boundary can change the size and frequency of the ELMs as the operating point may be at a different position on the stability space with respect to the boundary. To understand this better, we decipher the participating terms as follows: the ideal ballooning growth scales like:

$$\gamma \sim c_s/\sqrt{RL_p} \text{ where, } c_s = \sqrt{(T_e + T_i)/m_i} \qquad (1)$$

Such that $\gamma \propto m_i^{-1/2}$. Here $c_s$, $R$ and $L_p$ are the ion sound speed, major radius and the pressure gradient scale length respectively. Hence for a heavier ion, one should expect a lower growth rate. Now for high-n ballooning modes the diamagnetic frequency is $\omega_* \approx \omega_{*e} + \omega_{*i}$ with $\omega_{*e} >> \omega_{*I}$ as $T_e \sim T_i$, the electron gradients are often steeper, and the electron response controls the mode frequency. However, $\omega_{*i}$ could be important for higher $Z_{eff}$ as $\omega_{*i} = k_y T_i/(Z_{eff} eBL_{ni})$. So, the ion diamagnetic stabilization will be weaker with higher $Z_{eff}$. Now the diamagnetic stabilization criterion is roughly $\gamma/\omega_* \leq 1$. Even though, $\omega_{*i}$ has weaker contribution compared to $\omega_{*e}$, $\omega_{*i}$ can often be important for impurity seeding experiments as $\omega_{*i} \propto 1/Z_{eff}$. Nevertheless, the net stabilizing effect is stronger for heavier ions (due to weaker $\gamma$), which is why a fusion plasma using deuterium and tritium (D-T) can achieve higher performance than one using hydrogen (H) alone [48]. The presence of heavier impurity ions like B may potentially increase the average ion mass, but this can also be impacted negatively by diluting the fusion fuel and decreasing the ion diamagnetic drift via increased $Z_{eff}$. Some studies have shown that impurities can stabilize the pedestal by reducing the electron temperature and pressure gradients, but these results can be complex [49,50].

Since there are several differences in the plasma parameters like $n_e^{ped}$, $T_e^{ped}$, $Z_{eff}$ etc. among the reference discharge and the discharge(s) with B (refer to Fig. 4), we now scan the pedestal heights, widths and $Z_{eff}$ of the high B case (#179873) to match with those of the reference case (#179863). The idea is to match the high B case to the reference case step by step with one parameter or attribute (like height or width) of a parameter being matched progressively over the scans. This exercise may reveal the role of the pedestal profiles and $Z_{eff}$ for the decoupling of the PB boundaries. All the scans are done in the VARYPED module [51] starting with the kinetic EFIT [52], while keeping the total stored energy constant. As a first set of scans, either the pedestal heights of $n_e^{ped}$ or that of $T_e^{ped}$, or both, for the high B case are adjusted in such a way that they now match with the reference case. When the heights of both $n_e^{ped}$ and $T_e^{ped}$ are adjusted simultaneously (Blue trace in Fig. 6(a)), the decoupling decreased and the PB stability boundary closed off.

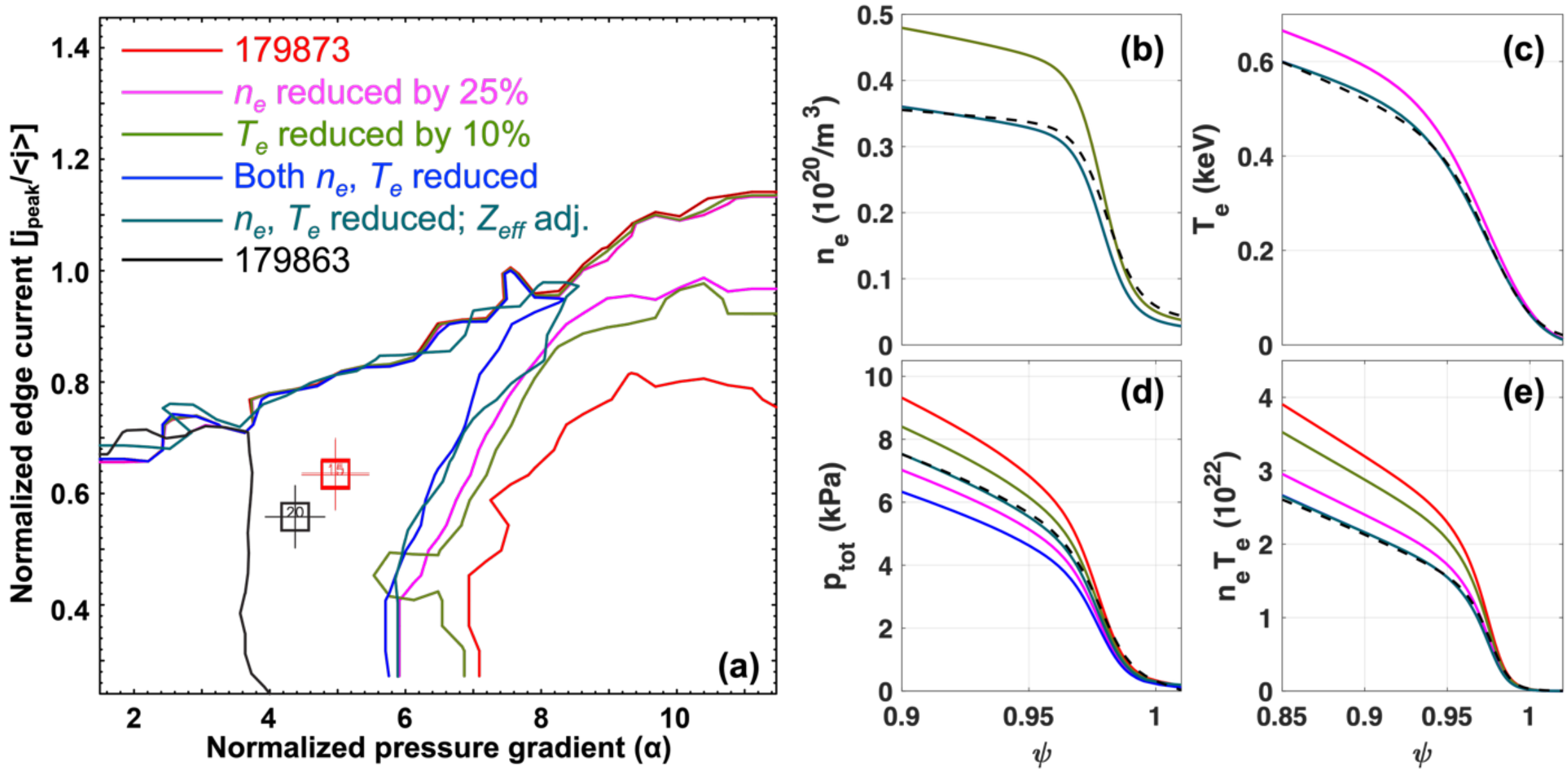

*Fig. 6: (a) PB stability of different cases in the profile height scan; red represents the original high B case (#179873); magenta represents the 25% decreased $n_e$, green represents 10% decreased $T_e$ and blue represents both $n_e$* and *$T_e$ decreased simultaneously; $Z_{eff}$ adjustment is shown in teal; (b-e): The profiles of $n_e$, $T_e$, $p_{tot}$ and $n_eT_e$ are shown on the right along with the reference (#179863; dashed) for comparison. All the traces are not visible in (b-c) due to overlaps.*

Next, we adjusted the $Z_{eff}$ by reducing the C and B densities in such a way that the C and B densities now match those of the reference case. In the process of matching C and B densities, $Z_{eff}$ also now matches with that of the reference case. The stability boundaries still show a closed stable boundary (teal), which is not significantly different from the pedestal height adjustments. The profiles of $n_e$, $T_e$, $n_eT_e$ and $p_{tot}$ are shown on the right for the different cases in the scan.

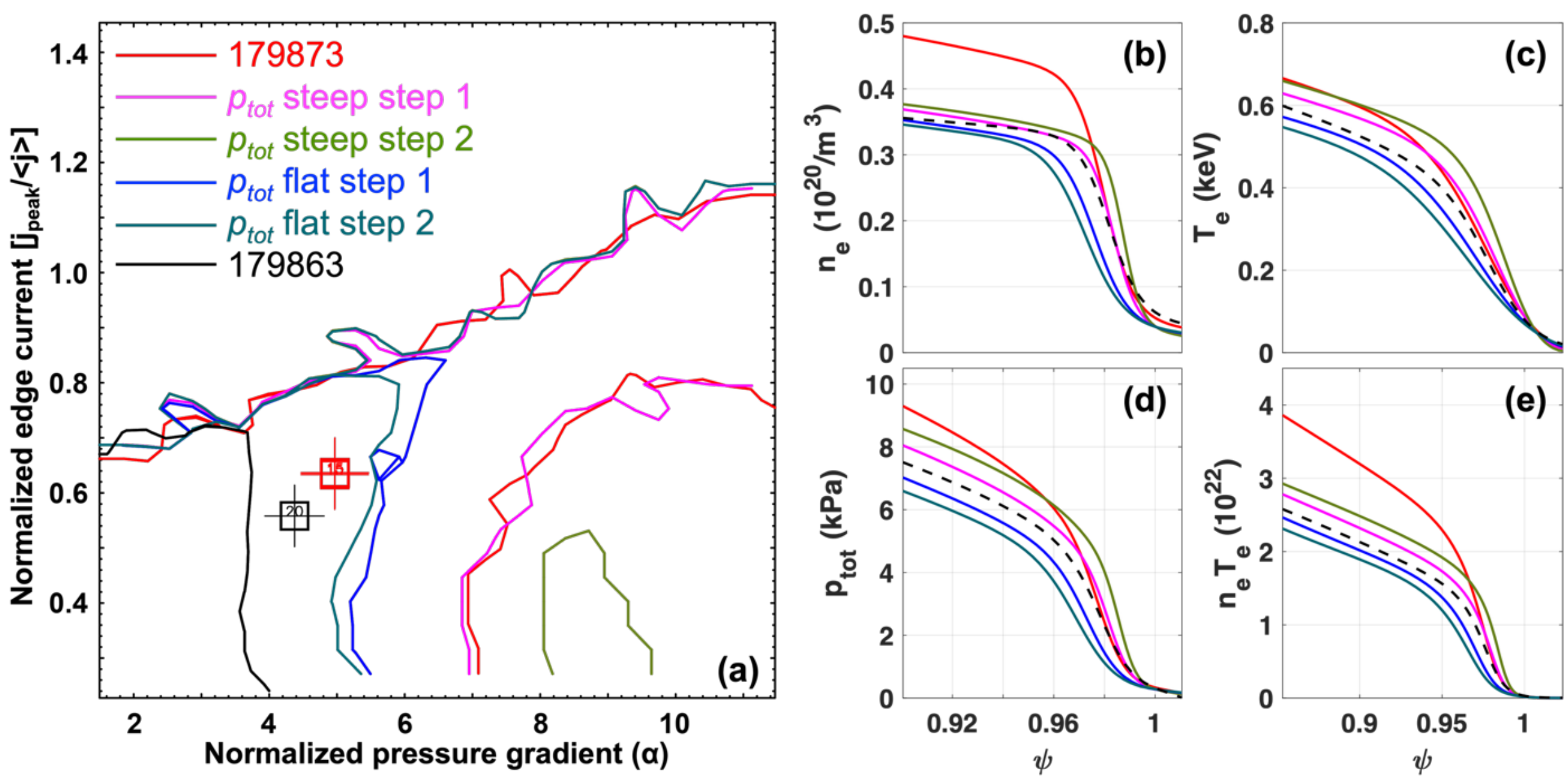

*Fig. 7: (a): PB stability of different cases in the profile flatness scan; Red represents the original high B case (#179873); magenta and green represent the progressive profile steepening while blue and teal represent progressive flattening respectively. Magenta, green, blue and teal are created from the simultaneously adjusted $n_e$, $T_e$, $Z_{eff}$ values of the original case to match with the reference case; (b-e): The profiles of $n_e$, $T_e$, $p_{tot}$ and $n_eT_e$ are shown on the right along with the reference (#179863; dashed) for comparison.*

Finally, the gradient of $p_{tot}$ could be another important factor that can affect the PB stability [16]. To illustrate this effect a set of equilibria were generated in which the profiles in the pedestal region were rigidly shifted radially starting from adjusted pedestal case from the last set of scans where the pedestal heights of both $n_e^{ped}$, $T_e^{ped}$ as well as $Z_{eff}$of the high B case are matched with those of the reference case. It is observed that rigid outward shift of the pressure profile (steepening of $p_{tot}$ profile) increases the decoupling, while inward shift (profile flattening) closes the PB boundaries (Fig. 7(a)). The main difference is observed in the ballooning stability with negligible changes in the peeling stability. Essentially, the ballooning stability improves with pressure profile steepening and vice versa. This observation is in contrast with what was observed with Li [16]. With Li, ballooning stability improves and peeling stability is reduced as the extent of the region with reduced gradient near the separatrix expands [Fig. 19 of 16]. Note that there is no trend of inward shift of the maximum gradient region with increasing B in our experiments as well. Only an increased pedestal and hence an inward propagation of the pedestal top to accommodate that increase is observed (refer to Fig. 4).

These scans show that the decoupling of the PB boundaries is not due to the shift of the maximum gradient region, but most likely is due to the increase in $n_e^{ped}$ and $T_e^{ped}$ pedestal height and may be because of the increase in $Z_{eff}$too via changes in the diamagnetic stabilization. Finally, adverse effect on the diamagnetic stabilization via increased $Z_{eff}$due to B injection and thereby fuel dilution, is not apparent from the ELITE scans (blue and teal traces of Fig. 6(a)). Note that high collisionality, in principle, can reduce the effectiveness of diamagnetic stabilization. In this case, collisionality kept on increasing with B injection rate, as evident from the significant increase in $n_e$ while modest change in $T_e$ is observed (refer to Fig. 4). Even in that case with increasing collisionality, diamagnetic stabilization seems to be improving, contrary to the standard understanding. This suggests that there could be a significant role of the turbulence scenario at the maximum gradient region of the pedestal, in determining the MHD growth rate and hence the decoupling of the PB boundaries, as observed earlier in DIII-D [20]. To address this issue, a thorough investigation of the turbulence scenario at the pedestal is essential.

*2.5.Increased turbulence levels might lead towards enhanced turbulence driven transport*

Once it is apparent that the pedestal stability is altered substantially by the injection of impurities, it is worthwhile to explore the role of turbulence in achieving the difference in stability. Turbulence analysis will help us to understand the reason behind the variation in pedestal recovery time and $f_{ELM}$ in the three representative cases. Let us start to investigate the turbulence scenario with the intermediate wavenumber density fluctuations ($\tilde{n}$) measured by the Doppler backscattering (DBS) diagnostic [53]. DBS diagnostics is used to measure $\tilde{n}$ in the pedestal in this discharge series with varied amounts of injected B. The measured $\tilde{n}$ has $k_\theta \rho_s \sim 0.4 - 1.2$ where $k_\theta$ is the perpendicular fluctuation wavenumber and $\rho_s$ is the ion Larmour radius assuming $T_e \cong T_i$. The RMS values of $\tilde{n}$ is estimated by integrating the turbulence intensity over the Doppler shifted part of the spectrum.

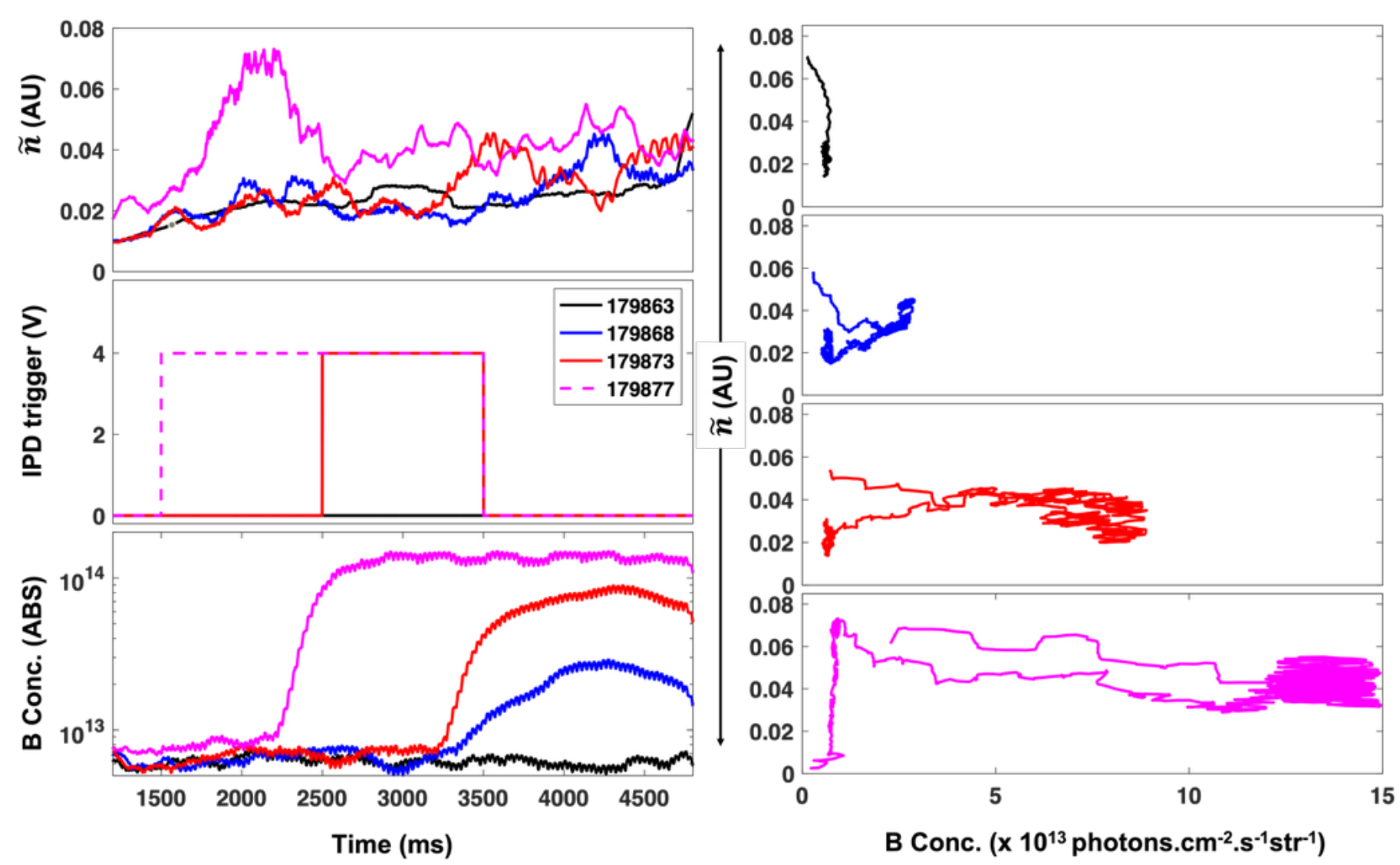

*Fig. 8: (Left) Time evolution of RMS $\tilde{n}$ at $\rho$ = 0.97, the voltage signal to the IPD and the concentration of B in the SPRED spectrometer for the four representative discharges 179863, 179868, 179873 and 179877. (Right): Hysteresis observed when RMS $\tilde{n}$ is plotted as a function of B concentration.*

Fig. 8 (left panel) shows the time evolution of $\tilde{n}$, the voltage signal to the IPD and the level of B (in absolute (ABS) units of photons.cm$^{-2}$.s$^{-1}$str$^{-1}$) in the SPRED spectrometer [39] for four of the representative discharges we are discussing so far. #179863 being the reference discharge shows no B in the SPRED signal. Note that the base level of B intensity in this discharge is ~$6 \times 10^{12}$ (ABS). When the B level increases above a threshold of ~$1.5 \times 10^{13}$ (ABS) in the medium B discharge (#179868), time-averaged $\tilde{n}$ RMS value increases with B level and as the B level drops $\tilde{n}$ decreases (at ~4000 ms) but stays higher even when B drops below the threshold, indicating a hysteresis. The time averaging for $\tilde{n}$ RMS is done over a running time window of 244.5 ms. When the B level was increased further in discharge #179873, time-averaged $\tilde{n}$ kept on increasing even when the B intensity goes beyond the threshold value of $1.5 \times 10^{13}$ (ABS). But with a substantial amount of B, showing B intensity of $\sim 4 \times 10^{13}$ (ABS) in SPRED at 3500 ms, triggers a rapid decrease in $\tilde{n}$. The time-averaged $\tilde{n}$ RMS value stays low for a while and again starts to increase once the B level starts to drop at 4300 ms, even though the threshold B level of $1.5 - 2 \times 10^{13}$ (ABS) has not been obtained yet, indicating a strong hysteresis again as we saw in #179868. In #179877, the B level before ~2000 ms was already higher than the reference discharge due to B accumulation from previous discharges and time-averaged $\tilde{n}$ RMS starts increasing even before B reaches the plasma, but once new B reaches the plasma, B level increased rapidly to a value which is almost twice the maximum B intensity in the high B shot (#179873). Consequently, $\tilde{n}$ drops to a slightly lower level and stays low throughout the shot as B level never drops again in the discharge.

The observed hysteresis is plotted in Fig. 8 (right panel). The hysteresis loop in the evolution of $\tilde{n}$ in response to the B injection rate widens with B concentration in the core and indicates the establishment of a feedback loop between turbulence, particle transport, and the resulting modification of pedestal conditions, such as pedestal height and gradients. Importantly, this behavior persists despite a reduced particle source due to improved wall conditioning, ruling out recycling-driven explanations. Instead, the hysteresis points to the emergence of a regulated transport channel, in which impurity-driven turbulence is consistent with replacing intermittent ELM losses, which are rather instantaneous, with a continuous, turbulence-mediated particle exhaust.

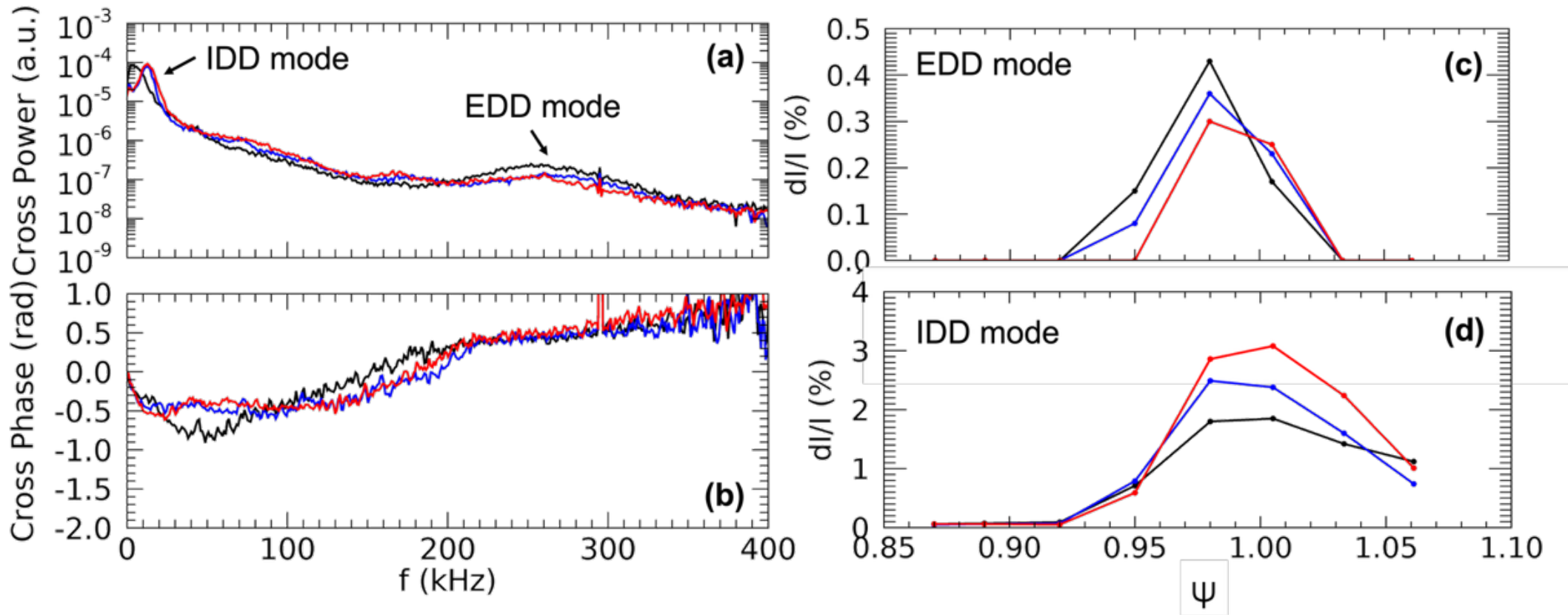

*Fig. 9: (a-b) cross power and cross phase of two consecutive BES channels at the maximum gradient region of the pedestal for the three representative discharges; (c-d) Radial profiles of the EDD and IDD modes show that both these modes are peaking at the pedestal maximum gradient region. Black is the reference (#179863), blue is low B (#179868) and red is the high B (#179873) discharge respectively.*

Let us now shift our focus towards the turbulence analysis with low wavenumber ($k_\theta \rho_s \sim 0.01$-$0.03$) local density fluctuations measured by the beam emission spectroscopy (BES) diagnostic [54]. Fig. 9(a)-(b) show the cross power and cross phase of two consecutive BES channels at the maximum gradient region of the pedestal. Dual modes are seen in the pedestal maximum gradient region with different trends of amplitudes for the ion diamagnetic direction (IDD) and the electron diamagnetic direction (EDD) mode with the difference in B concentration. Here, the positive and negative phases represent the EDD and IDD respectively. Fig. (c)-(d) shows that both these modes are peaking at ~ $\psi_N = 0.98$. The high frequency (~260 kHz) EDD mode decreases while the low frequency (<10 kHz) IDD mode increases with the increase in B injection.

To isolate the effect of these two modes in the inter-ELM pedestal recovery and transport, we filtered out the specific frequency regions of these modes from the raw BES (fast) signal for the discharge #179877, showing long ELM-free periods, such that the effects become most visible. Fig. 10(a) and 10(b) show the raw signal of channel #37 (at $\psi_N \sim 0.98$) in blue and the bandpass filtered signal in red. Fig. 10(a) shows filtered signal at 2-20 kHz representing the IDD mode while 10(b) shows the filtered signal at 190-280 kHz representing the EDD mode respectively. Apparently, the IDD mode represents almost entirety of the fast BES signal with negligible contribution from the EDD mode. This is further demonstrated in Fig. 10(c) and 10(d) where the IDD filtered signal and the spectrogram show the inter-ELM dynamics of the BES signal being represented by the IDD mode (in yellow). Finally, in 10(e) we expanded one of the ELM-free periods (3000-3200 ms) to illustrate the evolution of the IDD mode. It seems that the IDD mode has a dual character with the manifestation of quasi-coherent as well as bursty evolution in the inter-ELM period. Further, in 10(f), the baseline of the Dα is plotted to look for any correlation between the IDD bursts and the Dα baseline. It is evident that they are correlated as shown by the vertical broken black arrows. The mode fluctuation envelope in black shows increased fluctuations corresponding to enhanced Dα baseline, suggesting that the IDD is responsible for majority of the transport, especially particle transport (as Dα is a stronger function of density at the edge [55]), in the inter-ELM period. While direct quantification of particle flux is beyond the scope of this work, the consistency between fluctuation measurements and particle exhaust proxies, like the Dα baseline, strongly supports a turbulence-driven transport mechanism. It can be noted here that the increase in pedestal density observed with B injection does not contradict the presence of enhanced particle transport; rather, it reflects the removal of intermittent ELM particle losses (with lower $f_{ELM}$ or achieving ELM-free periods) and the establishment of a regulated pedestal in which impurity-driven turbulence provides continuous inter-ELM particle exhaust.

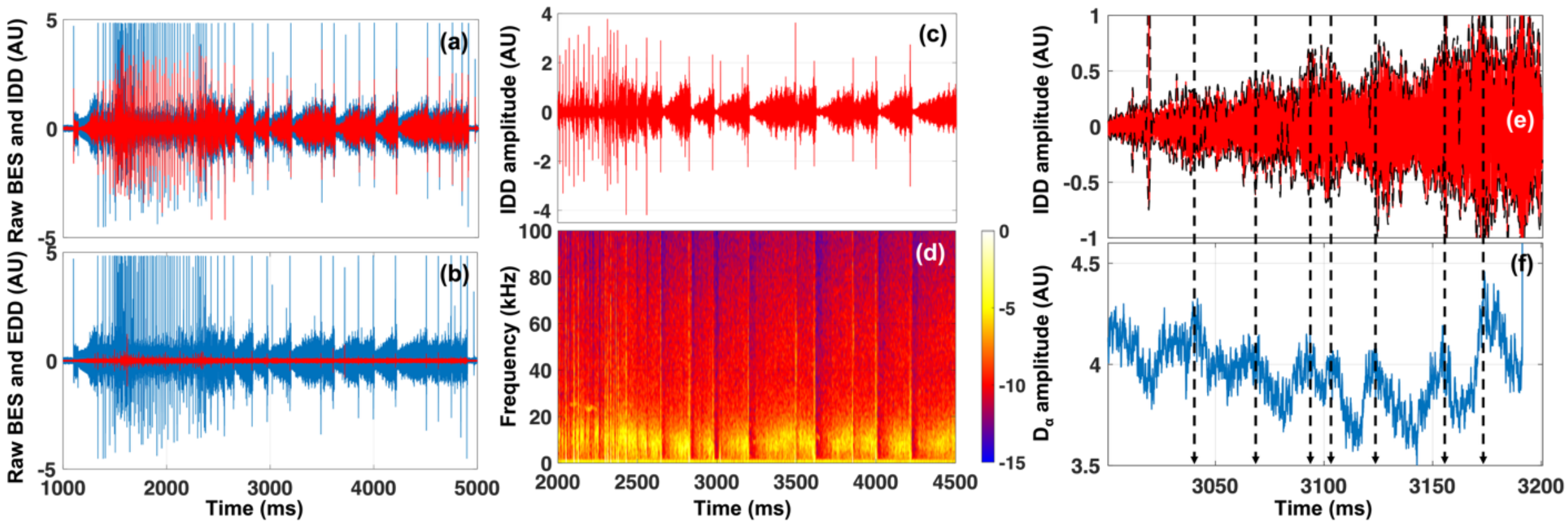

*Fig. 10: (a-b): superimposed raw (in blue) and frequency filtered signals (red) corresponding to the IDD and EDD modes; (a): frequency filter is applied in 2-20 kHz for the IDD mode while (b) frequency filter is applied in 190-280 kHz for the EDD mode; (c) frequency filtered raw signal to*

*show the evolution of the IDD mode; (d) Cross-power spectrogram of the fast BES signal to show the dynamics in the ELM-free periods and the modulation of the IDD mode at $\psi_N \sim 0.98$; (e-f): stronger bursts of the IDD mode correspond to the increased Dα baseline during an ELM-free period at 3000-3200 ms, shown by vertical broken lines. Enhanced IDD mode observed with B and its correlation with the elevated inter-ELM Dα baseline shown in panels (e–f), is indicative of enhanced convective particle transport rather than broadband thermal diffusion.*

Magnetics show similar reduction in broadband fluctuations at ~300 kHz with B injection like the observation of the EDD mode (not plotted here for brevity). The rise of the IDD mode and being responsible for most of the inter-ELM transport, especially particle transport, is in contrast with the findings with Li injection [16]. With Li injection an EDD mode with a bursty-chirpy character prevailed and dominated the transport scenario in the inter-ELM period. However, in LHD, a similar behavior of the EDD and the IDD was observed with B injection [56]. An analytical estimate of the transport characteristics is attempted in the next sub-section.

*2.6.Enhanced transport in the pedestal is the key*

It has been shown in the last sub-section that the low frequency IDD mode is responsible for enhanced particle transport in the inter-ELM period. To further elucidate the role of pedestal turbulence in achieving the ELM-free operation or determining $f_{ELM}$, we would like to refer to the analytical power balance equations. The effect of turbulence driven transport in the power balance is shown nicely by equations (7) and (8) in [25,40]. Here we can invoke these equations [40], which says:

$$P_{\mathrm{sep}} = P_{\mathrm{ELM}} + P_{\mathrm{transport}} \qquad (P_{\mathrm{ELM}} = f_{\mathrm{ELM}}\Delta W_{\mathrm{ELM}}) \qquad (2)$$

$$\Gamma_{\mathrm{sep}} = \Gamma_{\mathrm{ELM}} + \Gamma_{\mathrm{transport}} \qquad (\Gamma_{\mathrm{ELM}} = f_{\mathrm{ELM}}\Delta N_{\mathrm{ELM}}) \qquad (3)$$

where $P_{\mathrm{sep}}$ and $\Gamma_{\mathrm{sep}}$ denote the power and particle flux fed to the plasma across the separatrix and the subscripts '*ELM'* and '*transport*' denote the perpendicular energy and particle losses due to the ELM event itself and transport in the inter-ELM period, respectively. Thus, $P_{\mathrm{sep}}$ is the sum of the Ohmic power ($P_{\mathrm{OH}}$) and the auxiliary heating power, which in this case is $P_{\mathrm{NBI}}$ (refer to Fig. 1(a)) minus the radiated power from the confined plasma $P_{\mathrm{rad}}$ [57]. $P_{\mathrm{rad}}$ is the same and constant for all these discharges in the shot series at 1.2 MW in the stationary phase. With $P_{\mathrm{OH}}$ of 0.23 and $P_{\mathrm{NBI}}$ of 1.96 MW for all the discharges, including the reference and low $f_{ELM}$ (or high B) cases, $P_{\mathrm{sep}}$ is 0.99 MW for all these discharges. Note that here $P_{\mathrm{ELM}}$ (= $f_{\mathrm{ELM}}\Delta W_{\mathrm{ELM}}$) is 0.56 and 0.81 MW for the high B (#179873) and the reference (#179863) cases respectively. Thus, for $P_{\mathrm{sep}}$ with a variation of < 2%, $P_{\mathrm{transport}}$ must be different for these two cases to account for the ~ 45% variation in $P_{\mathrm{ELM}}$. Precisely speaking, $P_{\mathrm{transport}}$ must be higher for the high B discharge as compared to the reference discharge. As per the numbers given above, $P_{\mathrm{transport}}$ will be 0.43 and 0.18 MW for the high B and reference discharges respectively, rendering the transport driven power loss to be ~2.4 times higher in the high B case. Although the inferred $P_{\mathrm{transport}}$ increases with B injection even at moderate injection rates (viz. 4.5 mg/s), this increase is consistent with enhanced convective particle transport, which necessarily carries energy across the pedestal without implying a transition to strongly diffusive heat transport.

As stated earlier, particle balance is non-trivial due to change in wall recycling following boronization. Additionally, injection of varying amount of low-Z impurity itself, makes the scenario even more difficult to interpret from the viewpoint of particle source. That is the reason behind not performing traditional transport studies using TRANSP [58] power balance calculations. There is a 40% reduction in the Dα baseline for the B injected discharges as compared to the reference discharge. This is due to the change in wall recycling as observed earlier [59-61]. However, despite the modulation of the Dα baseline correlating with the stronger bursts of the IDD mode, indicating enhanced particle transport with B, a quantitative estimation of $\Gamma_{ELM}$ and $\Gamma_{transport}$ is difficult to obtain from the experimental data. A modelling initiative will be taken in near future to investigate this observation further.

## 3. Discussion

Our results demonstrate achievement of long ELM-free periods of operation with B injection for the first time in the DIII-D tokamak. This experiment has also demonstrated for the first time enhanced ballooning stability with B injection leading towards wide decoupling of the PB boundaries and thereby paving the way for easier super-H access in tokamak operation. These two-fold demonstrations can lead towards the solutions for two of the outstanding challenges of reactor grade operations simultaneously: (i) controlling and eventually eliminating the ELMs that can dump huge transient heat and particle loads on the plasma facing components, thereby limiting their lifetime or rendering the reactor completely useless and (ii) paving the way to achieve high confinement robustly, thereby achieving the critical values of the fusion triple product. However, it can be noted that the ELM-free scenario demonstrated here ends with larger ELMs after each long ELM free period, resulting in enhanced heat and particle loss to the wall. A better control of the amount of impurity injection in future may lead to sustained ELM-free scenario.

Real time wall conditioning, using B injection is being developed and under consideration for ITER [62]. The added advantages of B injection, demonstrated here, can help further this effort. It can be noted here that the amount of B required to achieve the desired results, as demonstrated in this experiment, is practically half of the amount required for Li [16].

The path towards lowering of $f_{ELM}$ and thereby achievement of long ELM-free periods is clearly linked to the selective excitation of turbulence modes at the pedestal and facilitating inter-ELM turbulence driven transport. An IDD mode is observed in this case to increase inter-ELM particle transport and thus helping the pedestal density and hence pressure to go up with minimal change of temperature. The strong temporal correlation between the growth of low-frequency ion-direction density fluctuations and the rise in the inter-ELM Dα baseline indicates that impurity injection enhances particle transport through turbulence-mediated mechanisms, rather than through increased MHD activity or radiative losses. Further, the net diamagnetic stabilization of high-n ballooning modes is, most probably, achieved by reduced MHD mode growth rates due to heavier ion injection. The increase in $Z_{eff}$, even though adversely affecting the ion diamagnetic stabilization, seems to be less detrimental while determining the pedestal stability boundaries for ELMs to occur. Thus, this real time wall conditioning technique with low-Z impurity, which might be feasible to implement in reactor conditions, if navigated perfectly, can offer immense benefits for the reactor grade operation scenarios in future fusion reactors.

## 4. Methods

### *4.1.DIII-D tokamak*

DIII-D [37] National Fusion Facility is a large aspect ratio tokamak research device operated by General Atomics in San Diego, California, for the U.S Department of Energy. The DIII-D program is pioneering the ways to establish the scientific basis for the optimization of the tokamak approach to nuclear fusion which is dubbed as the cleanest and safest energy source of the next generation. This experiment is performed with type-I ELMing H-mode plasmas in the lower single null (LSN) scenario in DIII-D [37], heated with 2 MW NBI (injected power, $P_{inj}$, shown in Fig. (a)). Plasma discharges in this experiment were designed to study the effect of real time wall conditioning with B powder injection on the plasma conditions. Plasma current $I_p$, toroidal magnetic field $B_T$ and hence $q_{95}$ are kept constant is these discharges at 1 MA, 2T and 5 respectively. The normalized plasma pressure, $\beta_N$ (= $\beta a B_T/I_p$, $a$ being the minor radius and $\beta$ being the ratio of the plasma pressure to the magnetic pressure, expressed in percentage) is $\sim$ 1.5 in all the discharges of this series. These discharges are operated with a total input co-current torque of 2.1 Nm (Fig. (b)).

### *4.2.Diagnostics for kinetic profile and fluctuation measurements*

A multi-chord, multi-pulse Thomson scattering (TS) system is used to measure electron density ($n_e$) and electron temperature ($T_e$) profiles at a high spatial and time resolution, along a vertical chord in the machine. [38]. The pedestal height and width for both $n_e$ and electron temperature $T_e$ are extracted using a modified hyperbolic tangent (tanh) fit applied to the pedestal and near-pedestal regions [63]. This fitting function comprises a tanh component to represent the pedestal, a linear function with finite slope for the region inboard of the pedestal, and a constant term to describe the region outboard of the pedestal. As described in Ref. [63], the pedestal height is defined as the sum of the background offset given by the constant term and the vertical amplitude of the tanh function. The pedestal width is obtained directly from a fit parameter and corresponds to the spatial extent of the region with steep gradients in the profile. Uncertainties in the pedestal height and width are determined through standard error propagation of the individual TS measurements. The statistical uncertainties of each TS measurement are derived from the one-standard-deviation noise contributions associated with electronics, photon statistics, and detector response [38]. Owing to the system's high signal throughput, low background levels, rigorous calibration procedures, and the use of multiple TS profiles in the fitting process, the resulting uncertainties in the fitted pedestal parameters are generally small. Toroidal rotation and ion temperature are obtained from the charge exchange recombination (CER) measurements [64 and the references therein].

Doppler backscattering (DBS) diagnostic [53] is used to measure intermediate wavenumber density fluctuations ($\tilde{n}$). DBS diagnostics is used to measure $\tilde{n}$ in the pedestal in this discharge series with varied amounts of injected B. The measured $\tilde{n}$ has $k_\theta \rho_s \sim 0.4 - 1.2$ where $k_\theta$ is the perpendicular fluctuation wavenumber and $\rho_s$ is the ion Larmour radius assuming $T_e \cong T_i$. The RMS values of $\tilde{n}$ is estimated by integrating the turbulence intensity over the Doppler shifted part of the spectrum. Low wavenumber ($k_\theta \rho_s$ ~ 0.01-0.03) local density fluctuations are measured by the beam emission spectroscopy (BES) diagnostic. Cross phase of two consecutive BES channels at the maximum gradient region of the pedestal is calculated to see the propagation directions of the IDD and EDD.

### *4.3.Kinetic equilibria reconstruction*

By using the PyD3D analysis toolkit [65], a multi-step workflow is employed to incorporate pressure and plasma current density constraints into the equilibrium reconstruction, thereby improving the accuracy of the reconstructed equilibrium. Each iteration of the workflow consists of three main steps: (1) profile fitting based on an existing equilibrium; (2) power balance calculations to determine the total pressure and plasma current components; and (3) equilibrium reconstruction incorporating pressure and current density constraints. In practice, two to three iterations are typically sufficient to obtain high-quality equilibria suitable for transport/pedestal stability studies. Here both C and B are considered as impurities. In the power balance step, NUBEAM [66] is used to calculate the NBI-driven current and fast-ion pressure, while the bootstrap current is evaluated from the measured pressure profiles using the Sauter model [67]. The ONETWO [68] code is then applied to combine the thermal and fast-ion pressure contributions to obtain the total pressure, and to compute the total plasma current density by accounting for externally driven current, bootstrap current, and Ohmic current determined from poloidal field diffusion. The equilibrium reconstruction step is carried out using the EFIT [69] code.

### *4.4.ELITE calculations*

ELITE is a single fluid MHD stability (eigenvalue) code. It includes diamagnetic stabilization via analytic models or fits to two fluid calculations. The stability boundary is determined by the ratio of growth rate to diamagnetic stabilization term on the grid of the normalized pedestal pressure gradient ($\alpha \propto dp^{PED}/d\rho$, $\rho$ being the normalized plasma radius) and normalized pedestal current density ($j_N$), and then tracing the contour for which this ratio is 1 [16,25,42].

**Acknowledgements**

This material is based upon work supported by the U.S. Department of Energy, Office of Science, Office of Fusion Energy Sciences, using the DIII-D National Fusion Facility, a DOE Office of Science user facility, under Awards No. DE-SC0019352, DE-FG02-08ER54999, DEAC02-09CH11466, DE-AC52-07NA27344, DE-SC0019004 and DE-FC02-04ER54698. This report was prepared as an account of work sponsored by an agency of the United States Government. Neither the United States Government nor any agency thereof, nor any of their employees, makes any warranty, express or implied, or assumes any legal liability or responsibility for the accuracy, completeness, or usefulness of any information, apparatus, product, or process disclosed, or represents that its use would not infringe privately owned rights. Reference herein to any specific commercial product, process, or service by trade name, trademark, manufacturer, or otherwise does not necessarily constitute or imply its endorsement, recommendation, or favoring by the United States Government or any agency thereof. The views and opinions of authors expressed herein do not necessarily state or reflect those of the United States Government or any agency thereof.